\documentclass[twocolumn,aps,superscriptaddress]{revtex4-1}
\usepackage[T1]{fontenc}
\usepackage[latin9]{inputenc}

\usepackage{amsmath}
\usepackage{amssymb}
\usepackage{bbm}
\usepackage{braket}
\usepackage{ulem}
\allowdisplaybreaks

\usepackage{graphicx}
\usepackage[colorlinks=true,citecolor=blue,linkcolor=blue]{hyperref}
\usepackage{color}

\newcommand{\equref}[1]{Eq.~(\ref{#1})}
\newcommand{\equsref}[2]{Eqs.~(\ref{#1}) and (\ref{#2})}

\newcommand{\figref}[1]{Fig.~\ref{#1}}

\newcommand{\refcite}[1]{Ref.~\onlinecite{#1}}
\newcommand{\refscite}[1]{Refs.~\onlinecite{#1}}

\newcommand{\diff}{\mathrm{d}}

\newcommand{\pdagger}{{\phantom{\dagger}}}

\renewcommand{\approx}{\simeq}

\renewcommand{\vec}[1]{\boldsymbol{#1}}

\newcommand{\change}[1]{{#1}}
\usepackage{ulem}
\renewcommand{\emph}[1]{\textit{#1}}

\makeatother

\begin{document}

\title{\change{Limits on dynamically generated spin-orbit coupling: Absence of $l=1$ Pomeranchuk instabilities in metals}}

\author{Egor I. Kiselev}

\affiliation{Institut für Theorie der Kondensierten Materie, Karlsruher Institut
für Technologie, 76131 Karlsruhe, Germany}

\author{Mathias S. Scheurer}

\affiliation{Institut für Theorie der Kondensierten Materie, Karlsruher Institut
für Technologie, 76131 Karlsruhe, Germany}

\author{Peter Wölfle}

\affiliation{Institut für Theorie der Kondensierten Materie, Karlsruher Institut
für Technologie, 76131 Karlsruhe, Germany}

\affiliation{Institut für Nanotechnologie, Karlsruher Institut für Technologie,
76344 Karlsruhe, Germany}

\author{Jörg Schmalian}

\affiliation{Institut für Theorie der Kondensierten Materie, Karlsruher Institut
für Technologie, 76131 Karlsruhe, Germany}

\affiliation{Institut für Festkörperphysik, Karlsruher Institut für Technologie,
76344 Karlsruhe, Germany}

\begin{abstract}
An ordered state in the spin sector that breaks parity without breaking time-reversal symmetry, i.e., that can be considered as dynamically generated spin-orbit coupling, was proposed to explain puzzling observations in a range of different systems. 
Here we derive severe restrictions for such a state that follow from a Ward identity related to spin conservation. It is shown that $l=1$ spin-Pomeranchuk instabilities are not possible in non-relativistic systems since the response of spin-current fluctuations is entirely incoherent and non-singular. This rules out relativistic spin-orbit coupling as an emergent low-energy phenomenon. We illustrate the exotic physical properties of the remaining higher angular momentum analogues of spin-orbit coupling and derive a geometric constraint for spin-orbit vectors in lattice systems. 
\end{abstract}

\maketitle
\section{Introduction}
Fermi liquid (FL) theory \cite{Landau57,Landau57b,Baym84,VollhardtWoelfle90}
forms the intellectual foundation of our understanding of quantum
fluids such as $^{3}$He, fermionic atomic gases, as well as simple
metals and numerous strongly-correlated systems. The inherent instability
of the FL state was analyzed early-on by Pomeranchuk \cite{Pomeranchuk58},
who derived the threshold values $-\left(2l+1\right)$ for the phenomenological
Landau parameters $F_{l}^{s,a}$ beyond which newly ordered states
are expected to emerge. Here, $l$ corresponds to the angular momentum
channel under consideration and $s$ ($a$) refers to order in the
charge (spin) sector. These criteria were derived from an analysis
of the quasiparticle contribution to the energy of a FL.
Famous examples of Pomeranchuk instabilities are phase separation
for $F_{0}^{s}\rightarrow-1$, ferromagnetism for $F_{0}^{a}\rightarrow-1$,
or charge- and spin nematic order for $F_{2}^{s,a}\rightarrow-5$.
Our main interest here will be the behavior for $F_{1}^{a}\rightarrow-3$,
which was proposed by Wu and Zhang \cite{WuZhang04} and which would
lead to a state with order in the spin sector that breaks parity while
time-reversal symmetry remains intact. Such order has been invoked
to explain the behavior in systems as diverse as chromium \cite{Hirsh90a,Hirsh90b},
Sr$_{3}$Ru$_{2}$O$_{7}$ \cite{Yoshioka12}, the hidden order in
URu$_{2}$Si$_{2}$ \cite{Varma06}, or the physics in the vicinity
of a ferromagnetic quantum phase transition \cite{Chubukov09}. The
associated dynamic generation of spin-orbit coupling of \refcite{WuZhang04}
was analyzed in great detail in \refcite{Wu07}. Even $^{3}$He was
argued to be in the vicinity of a corresponding instability \cite{WuZhang04}.

In this paper, we demonstrate that $l=1$ Pomeranchuk
instabilities in the spin channel are not possible.
Specifically, the divergence in the quasiparticle susceptibility is
canceled by a vanishing vertex that connects the response
of quasiparticles and bare fermions. As a consequence, the response
of the system is entirely incoherent and the analysis of the energy
balance of \refcite{Pomeranchuk58} turns out to be incomplete. \change{The identical result is obtained within FL theory if the proper form of the spin current in terms of the quasiparticle distribution function is used.} 
The origin of this peculiar behavior is that the spin current, which is
closely related to the instability, is itself not a conserved quantity,
yet it is the current of a conserved density. This allows to draw
rigorous conclusions from associated Ward identities that exclude
second order phase transitions. At the level of the Hartree-Fock approximation,
this has been found earlier in \refcite{Quintanilla06}. Our arguments imply that the absence of a second-oder instability for $l=1$ is, in
fact, exact. Following an argument by Bloch (see \refscite{Bohm49,BrayAli09}),
we also show that a first order transition is not allowed either.
While some of those conclusions for Pomeranchuk instabilities could have been drawn from the vast
literature on the FL theory and its microscopic foundations
(see in particular the work by Leggett in \refcite{Leggett65}),
the ongoing discussion of this instability in \refscite{WuZhang04,Hirsh90a,Hirsh90b,Yoshioka12,Varma06,Chubukov09,Wu07} 
seems to warrant a detailed analysis of this issue. 
In addition, it is shown that relativistic spin-orbit coupling cannot emerge due to spontaneous symmetry breaking of the electron liquid at low energies.
Spontaneously generated spin-orbit interactions are only expected in higher angular momentum channels with unconventional residual symmetry groups exhibiting exotic physical behavior such as enhanced anomalous Hall conductivities. Finally, the implications of our findings for lattice systems are discussed. 

\section{Pomeranchuk instability}
\begin{figure}[tb]
\begin{center}
\includegraphics[width=0.9\linewidth]{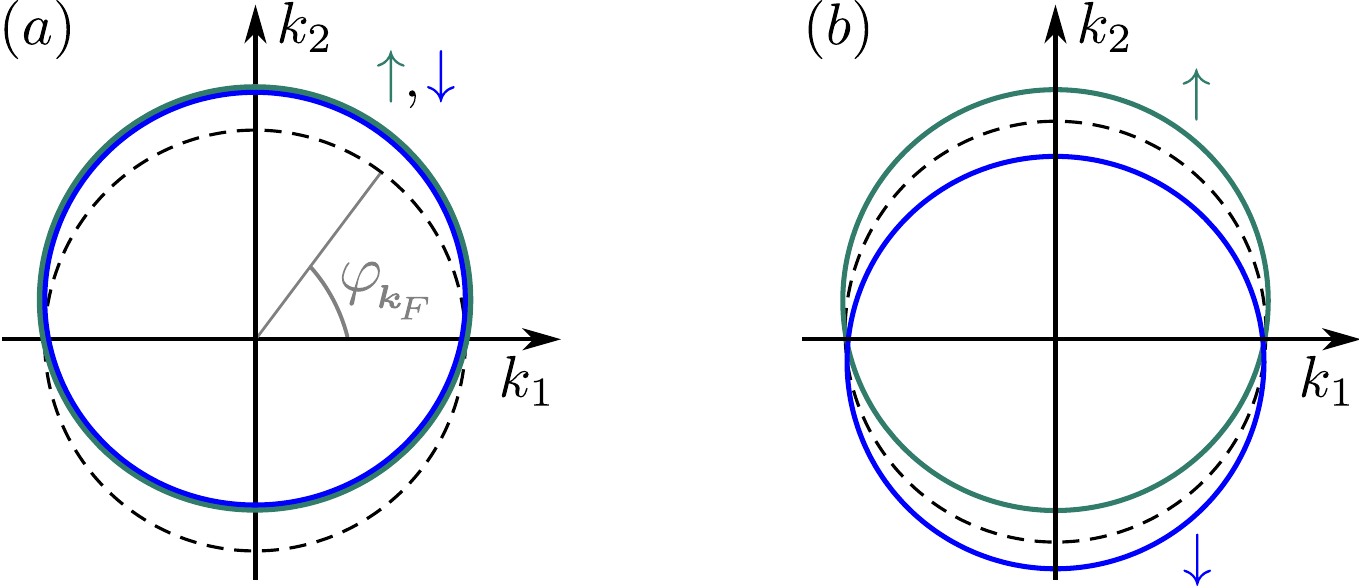}
\caption{Distortion of the spin-up (green) and spin-down (blue) Fermi surface for the $l=1$ Pomeranchuk instability in the (a) charge and (b) spin channel.}
\label{FermiSurfaceDeforms}
\end{center}
\end{figure}

The phenomenological formulation of FL theory is based
on the celebrated parametrization of the energy change due to quasiparticle
excitations \cite{Landau57,Landau57b} $\delta E_{{\rm qp}}=\sum_{\vec{k}\sigma}\varepsilon_{\vec{k}\sigma}^{*}\delta n^{\change{\text{qp}}}_{\vec{k}\sigma}$,
with single-particle energy 
\begin{equation}
\varepsilon_{\vec{k}\sigma}^{*}=v_F(|\vec{k}|-k_F) + \mu +\frac{1}{\rho_{F}}\sum_{\vec{k}',\sigma'}F_{\vec{k},\vec{k}^{\prime}}^{\sigma,\sigma'}\delta n^{\change{\text{qp}}}_{\vec{k}^{\prime}\sigma^{\prime}}, \label{EnergyOfQPs}
\end{equation}
where $\mu$, $k_F$, $v_{F}=\frac{m^{}}{m^*}v_F^0$, $\rho_{F}=\frac{m^{*}}{m}\rho_{F}^{0}$, and $m^{*}/m$ denote, respectively, the Fermi energy, momentum, velocity, the density of states, and the mass renormalization. 
For the interaction function we use the usual expansion for isotropic Fermi surfaces in the absence of spin-orbit coupling, $F_{\vec{k},\vec{k}^{\prime}}^{\sigma,\sigma'}=F_{\vec{k},\vec{k}^{\prime}}^{s}+\sigma\sigma^{\prime}F_{\vec{k},\vec{k}^{\prime}}^{a}$, $F^{r}(\theta)=\sum_{l=0}^{\infty}F_{l}^{r}P_{l}(\cos\theta)$,
with Legendre polynomials $P_l$ and $\cos\theta=\vec{e}_{\vec{k}}\cdot\vec{e}_{\vec{k}'}$.
Here $r=s$ and $r=a$ for the symmetric (charge) and anti-symmetric (spin)
channel, respectively. Pomeranchuk concluded that an instability with
a spontaneous deformation of the Fermi surface (see, e.g., \figref{FermiSurfaceDeforms})
\begin{equation}
k_{F}\rightarrow k_{F}+\delta k_{F,l}^{s}(\varphi_{\vec{k}_F})+\sigma\delta k_{F,l}^{a}(\varphi_{\vec{k}_F})
\end{equation}
occurs, when $F_{l}^{s,a}\rightarrow-\left(2l+1\right)$. Here $\varphi_{\vec{k}_F}$
is the angle of the Fermi wave vector relative to some arbitrarily
chosen axis. Specifically, the energy change due to quasiparticle excitations
was found to be \cite{Pomeranchuk58} 
\begin{equation}
\delta E_{{\rm qp}}=\frac{1}{\rho^0_{F}}\frac{m}{m^{*}}\hspace{-0.6em}\sum_{l,r=\left\{ s,a\right\} } \hspace{-1em}\left|\delta n_{l}^{{\change{\text{qp}}},r}\right|^{2}\left(1+\frac{F_{l}^{r}}{2l+1}\right)+{\cal O}\left(\left(\delta n^{\change{\text{qp}}}\right)^{4}\right).\label{eq:energyP}
\end{equation}
Here, $\delta n_{l}^{\change{\text{qp}},r}$ is the change in the quasiparticle occupation
in momentum space caused by $\delta k_{F,l}^{r}(\varphi_{\vec{k}_F})$
and acts as order parameter of the new phase.
If $m/m^{*}$ remains finite and the coefficient
of $\left|\delta n_{l}^{\change{\text{qp}},r}\right|^{2}$ becomes negative, an instability
to a state with $\delta n_{l}^{\change{\text{qp}},r}\neq0$ is energetically favored.
This happens once $F_{l}^{r}$ reaches the above threshold value.
The response of quasiparticles can also be characterized by the quasiparticle susceptibility \cite{WuZhang04}
\begin{equation}
\chi_{{\rm qp},l}^{r}=\frac{\rho_{F}}{1+\frac{F_{l}^{r}}{2l+1}}.\label{eq:susqp}
\end{equation}
For $l=0$ this corresponds to the well known expressions of the charge
and spin susceptibilities for the symmetric ($s$) and antisymmetric
($a$) channel, respectively. $\chi_{{\rm qp},l}^{r}$
diverges as $F_{l}^{r}\rightarrow-\left(2l+1\right)$, which coincides
with the Pomeranchuk instability criterion.

Below we demonstrate that the leading order expansion of the full
energy with respect to \change{the electron momentum distribution} $\delta n_{1}^{r}$ is rigorously given by 
\begin{equation}
\delta E_{l=1}^{r}=\frac{1}{\rho_{F}^{0}}\left|\delta n_{1}^{r}\right|^{2}+{\cal O}\left(\left(\delta n_{1}^{r}\right)^{4}\right),\label{eq:energy}
\end{equation}
ruling out the corresponding Pomeranchuk instability and demonstrating
that the analysis of the energy (\ref{eq:energyP}) of quasiparticles 
is not sufficient. This is closely related to the fact that the susceptibility
of the system is not adequately expressed in terms of its quasiparticle
contribution in \equref{eq:susqp}. 

In the microscopic foundation of FL theory \cite{AGB75},
one divides the single-particle Green's function $G_{\vec{k}}(\omega)$
in a coherent quasiparticle contribution and an incoherent background,
\begin{equation}
G_{\vec{k}}(\omega)=\frac{Z}{\omega+i0^{+}-\varepsilon_{\vec{k}}^{*}}+G^{{\rm inc.}}_{\vec{k}}(\omega).\label{eq:Gtot}
\end{equation}
$Z$ is the spectral weight and $\varepsilon_{\vec{k}}^{*}$ the
single-particle energy in \equref{EnergyOfQPs}. 
A detailed analysis of the susceptibilities of a many-body
system was given by Leggett \cite{Leggett65}, who finds 
\begin{equation}
\chi_{l}^{r}=\left(\gamma_{l}^{r}Z\right)^{2}\chi_{{\rm qp},l}^{r}+\chi_{{\rm inc},l}^{r}.\label{eq:sustot}
\end{equation}
Here, $\chi_{{\rm qp},l}^{r}$ is the quasiparticle response of
\equref{eq:susqp} and $\gamma_{l}^{r}$ the vertex that connects
the response of quasiparticles and bare fermions. Finally, $\chi_{{\rm inc},l}^{r}$
is the incoherent response of the system, a contribution that is directly
related to the incoherent part $G^{{\rm inc.}}_{\vec{k}}(\omega)$ of the single-particle Green's function (\ref{eq:Gtot}).
The incoherent response of the system can now be expressed as
\begin{equation}
\chi_{{\rm inc},l}^{r}=\chi_{l}^{r\left(\omega\right)}\equiv\lim_{\omega\rightarrow0}\lim_{\vec{q}\rightarrow\vec{0}}\chi_{l}^{r}(\vec{q},\omega),
\end{equation}
taking into account that $\chi_{{\rm qp},l}^{r}$ vanishes in this limit.

Next we discuss implications for susceptibilities that are caused
by conservation laws, i.e., associated with a Hermitian operator $\Phi$
that commutes with the Hamiltonian $H$ of the system, $[\Phi,H]=0.$ An example
is charge or particle-number conservation ($\Phi=N$). In a non-relativistic
system with Hamiltonian 
\begin{align}
\begin{split}
H & =  \int_{\vec{r}\alpha}\psi_{\alpha}^{\dagger}(\vec{r})\left(-\frac{\hbar^{2}\nabla^{2}}{2m}-\mu+U(\vec{r})\right)\psi^{\pdagger}_{\alpha}(\vec{r}) \\
 &  +  \frac{1}{2}\int_{\vec{r}\alpha,\vec{r}'\beta}\psi_{\alpha}^{\dagger}(\vec{r})\psi_{\beta}^{\dagger}(\vec{r}')V({\vec{r}-\vec{r}'})\psi^{\pdagger}_{\beta}(\vec{r}')\psi^{\pdagger}_{\alpha}(\vec{r}),\label{eq:Hamiltonian}
\end{split}\end{align}
the conservation of the components of the total spin $\Phi=S^{j}$
is another example. Here we use $\psi_{\alpha}$ ($\psi^\dagger_{\alpha}$) to represent the annihilation (creation) of a bare fermion of spin $\alpha$ and apply the convention $\int_{\vec{r}\alpha} \, \dots \, = \int \diff^{d}r\sum_{\alpha}\dots\,\,$.
Let us first focus on the case without crystal potential $U(\vec{r})$,
which yields the bare dispersion $\varepsilon_{\vec{k}}=\frac{\hbar^{2}k^{2}}{2m}-\mu$,
and comment on the implications of a finite crystal potential at the end. 

We write $\Phi=\int \diff^{d}r\rho^{\phi}(\vec{r})$
with density $\rho_{\vec{q}}^{\phi}=\frac{1}{V}\sum_{\vec{k},\alpha\beta}\psi_{\vec{k}+\frac{\vec{q}}{2}\alpha}^{\dagger}\phi_{\vec{k}}^{\alpha\beta}\psi^{\pdagger}_{\vec{k}-\frac{\vec{q}}{2}\beta}$ in momentum space 
and form factor $\phi_{\vec{k}}^{\alpha\beta}$. The charge density
corresponds to $\phi_{\vec{k}}^{\alpha\beta}=\delta_{\alpha\beta}$,
and a spin density amounts to $\phi_{\vec{k}}^{\alpha\beta}=\sigma_{\alpha\beta}^{j}$, $j=1,2,3$, with Pauli matrices $\sigma^j$. 
Let us assume that $\rho_{\vec{q}}^{\phi}$ commutes with the interacting (non-quadratic) part $H_{\text{int}}$ of the Hamiltonian,
\begin{equation}
 \left[\rho_{\vec{q}}^{\phi},H_{{\rm int}}\right]=0. \label{Prerequ}
\end{equation}
This is the case, e.g., for interactions of the form $H_{\text{int}} = f[\{\rho_{\vec{q}}^{\phi}\}]$ such as $H_{{\rm int}}=\sum_{\vec{q}}\change{V_{\vec{q}}}\rho_{\vec{q}}^{\phi}\rho_{-\vec{q}}^{\phi}$.
Most importantly, this also holds for both the spin and charge density in the case of the non-relativistic solid-state Hamiltonian (\ref{eq:Hamiltonian}) with electron-electron Coulomb
interaction. 
If \equref{Prerequ} applies, we can derive a Ward identity (see the Appendix \ref{Ward identity and its relation to susceptibilities}) that implies
\begin{equation}
\chi_{\rho}\left(\vec{q}=\vec{0},\omega\neq0\right)=0  \label{eq:dynsus}
\end{equation}
for the susceptibility $\chi_{\rho}$ of the conserved density $\rho_{\vec{q}}^{\phi}$ and
\begin{equation}
\chi_{J}^{\left(q\right)ij}\equiv\chi_{J}^{ij}\left(\vec{q}\rightarrow0,0\right)=-\sum_{\vec{k},\gamma\delta}\left(\phi_{\vec{k}}^{\gamma\delta}\right)^{2}\frac{\partial n_{\vec{k}}}{\partial k{}_{i}}\frac{\partial\varepsilon_{\vec{k}}}{\partial k_{j}},\label{eq:currsus}
\end{equation}
for the susceptibility $\chi_{J}$ of the current $\vec{J}_{\hspace{-0.1em}\phi}$ associated with $\rho_{\vec{q}}^{\phi}$.
In \equref{eq:currsus}, $n_{\vec{k}}$ denotes the momentum occupation and the limit $\vec{q}\rightarrow 0$  has to be performed along the $i$th direction after the limit $\omega\rightarrow0$,
which is indicated by the superscript $\left(q\right)$. Note, \equref{eq:currsus} is valid for an arbitrary dispersion $\varepsilon_{\vec{k}}$ and not limited to Galilean invariant systems.
The fact that the susceptibility of \equref{eq:currsus} is finite also excludes critical phases that might exist without a finite order parameter.

The implication of \equref{eq:dynsus} for susceptibilities of conserved
quantities is obvious and well established. Let the form factor be
$\phi_{\vec{k}}^{\alpha\beta}=\sigma^{j}_{\alpha\beta}$, $j=0,3$. Then follows from \equref{eq:dynsus}
that $\chi_{{\rm inc},0}^{s,a}=\chi_{0}^{s,a\left(\omega\right)}=0$,
i.e., the entire response of the system is coherent. The same Ward
identity can also be used to show that $\gamma_{0}^{s,a}=Z^{-1}$ 
and leads to the well known relation $\chi_{0}^{r}=\chi_{{\rm qp},0}^{r}$
with quasiparticle susceptibility given in \equref{eq:susqp}. This means that
the susceptibility of a conserved density is fully determined by the
quasiparticle response. Using the formalism of \refcite{Leggett65},
one can also determine the next order corrections in $q/\omega$,
\begin{equation}
\chi_{0}^{r}(\vec{q},\omega)=\frac{q^{2}}{\omega^{2}}\frac{m}{m^{*}}\left(1+\frac{F_{1}^{r}}{3}\right)+{\cal O}\left(\left(\frac{q}{\omega}\right)^{4}\right).\label{eq:chi0high freq}
\end{equation}

Let us next exploit the Ward identity (\ref{eq:currsus}) for the current.
If the Fermi energy is the largest energy scale, we can replace $\frac{\partial\varepsilon_{\vec{k}}}{\partial k_{i}}$
by $\frac{k_F}{m}\cos\varphi_{\vec{k}}$ in the current operator $\vec{J}_{\hspace{-0.1em}\phi}$ showing that the charge ($\phi_{\vec{k}}=\sigma^{0}$) and spin ($\phi_{\vec{k}}=\sigma^{3}$)
current susceptibilities correspond to $l=1$ instabilities with $r=s$ and $r=a$, respectively. 
Analyzing \equref{eq:currsus} yields our key result $\chi_{1}^{s,a}=\rho_{F}^{0}$, which, via Legendre
transformation, leads to \equref{eq:energy}. A Pomeranchuk instability
in the $l=1$ channel is therefore not possible. 

If one further uses the continuity equation $\frac{\omega^{2}}{q^{2}}\chi_{\rho}(\vec{q},\omega)=\chi_{J}(\vec{q},\omega)-\chi_{J}(\vec{q},0)$
one can identify the vertex and incoherent contribution for the spin-current
susceptibility \cite{Leggett65}
\begin{align}
\begin{split}
 \gamma_{1}^{r} & =  Z^{-1}\frac{m}{m^{*}}\left(1+\frac{1}{3}F_{1}^{r}\right), \\
 \chi_{{\rm inc},1}^{r} & =  \rho_{F}^{0}\left[1-\frac{m}{m^{*}}\left(1+\frac{1}{3}F_{1}^{r}\right)\right],\label{eq:susfin}\end{split}
\end{align}
where we used \equref{eq:chi0high freq}. 
Let us analyze \equref{eq:susfin} for the charge and spin response
separately. For a Galilean-invariant system, the charge current is
itself a conserved quantity and \equref{eq:dynsus} implies that
$\chi_{{\rm inc},1}^{s}=0$. Thus, we recover the celebrated result
$\frac{m^{*}}{m}=1+\frac{1}{3}F_{1}^{s}$. In addition, it follows
$\gamma_{1}^{s}=Z^{-1}$ and $\chi_{1}^{s}=\chi_{{\rm qp},1}^{s}=\rho_{F}^{0}$.
While no Pomeranchuk instability will take place in the $l=1$ charge
channel, these results are fully consistent with the analysis of \refcite{Pomeranchuk58}
as the entire response in the $l=1$ charge channel of Galilean-invariant
systems is coherent and captured by quasiparticle excitations.

The situation is different for currents that are, themselves, not
conserved quantities, such as the spin current. If one approaches
the Pomeranchuk threshold value, the vertex $\gamma_{1}^{a}$ vanishes
and there is no contribution of the susceptibility due to quasiparticles.
The divergence of the quasiparticle contribution of the susceptibility
is suppressed by the vanishing vertex. As a consequence, the entire
response becomes incoherent and the energetic analysis that led to
\equref{eq:energyP} is not applicable, leading to \equref{eq:energy}
instead. 

\change{It is interesting to note, however, that the correct result for the physical spin current susceptibility may be obtained within FL theory if the proper form of the spin current is used (see the Appendix \ref{Response functions within FL theory}). There we also derive the response of the spin (charge) momentum current density to an external field of $l=2$ symmetry, which is again very different from the $l=2$ quasiparticle susceptibilities of \equref{eq:susqp}, i.e., $\chi_2 \neq \chi_{\text{qp},2}$.}

The arguments given so far exclude critical behavior in the $l=1$ channel with diverging susceptibility. To exclude a state with finite order parameter $\delta n_{1}^{r}$ that might be reached via a first order transition, we first note that $\delta n_{1}^{a}$ ($\delta n_{1}^{s}$) is proportional to the expectation value of $\vec{J}_{\hspace{-0.1em}\phi}$ with $\phi = \sigma^3$ ($\phi = \sigma^0$).
We can then apply the arguments of \refscite{Bohm49,BrayAli09} to show that, given any state with finite expectation value of $\vec{J}_{\hspace{-0.1em}\phi}$, there is always a state with lower energy. Consequently, there is also no first order transition to an $l=1$ Pomeranchuk phase.

\section{Constraints on spontaneous generation of spin-orbit coupling}
The results presented above yield strong restrictions on the interaction-induced generation of spin-orbit coupling
\begin{equation}
 \Delta H = \int_{\vec{k},\alpha\beta} \psi^\dagger_{\vec{k}\alpha}\vec{g}(\vec{k}) \cdot \vec{\sigma}_{\alpha\beta}\psi^{\pdagger}_{\vec{k}\beta}, \quad \vec{g}(\vec{k}) = -\vec{g}(-\vec{k}), \label{OrderParameter}
\end{equation} 
with order parameter $\vec{g}(\vec{k})$.
Assume that the Hamiltonian (such as the non-relativistic Hamiltonian in \equref{eq:Hamiltonian} with $U=0$) has a symmetry group that is a direct product $\text{SO}(d)_L \otimes \text{SO}(3)_S$ in orbital and spin space with $d=3$ ($d=2$) for three-dimensional (two-dimensional) systems. We can then expand $\vec{g}(\vec{k})$ in terms of basis functions (spherical harmonics $Y_{l,m}$ for $d=3$ and $e^{\pm il\varphi_{\vec{k}}}$ for $d=2$) transforming under the irreducible representations of $\text{SO}(d)_L$. Noting that the basis functions of $l=1$ are superpositions of $\{k_j\}$, we conclude that the $l=1$ channel can be discarded. Since $l \otimes 1 = (l+1) \oplus l \oplus (l-1)$, \equref{OrderParameter} cannot contain a term that is invariant under $\text{SO}(d)_{L+S}$, the set of simultaneous spin and orbital rotations, which is the point group of a system with relativistic spin-orbit coupling. In this sense, relativistic spin-orbit coupling cannot occur as an emergent low-energy phenomenon.  

\begin{figure}[tb]
\begin{center}
\includegraphics[width=\linewidth]{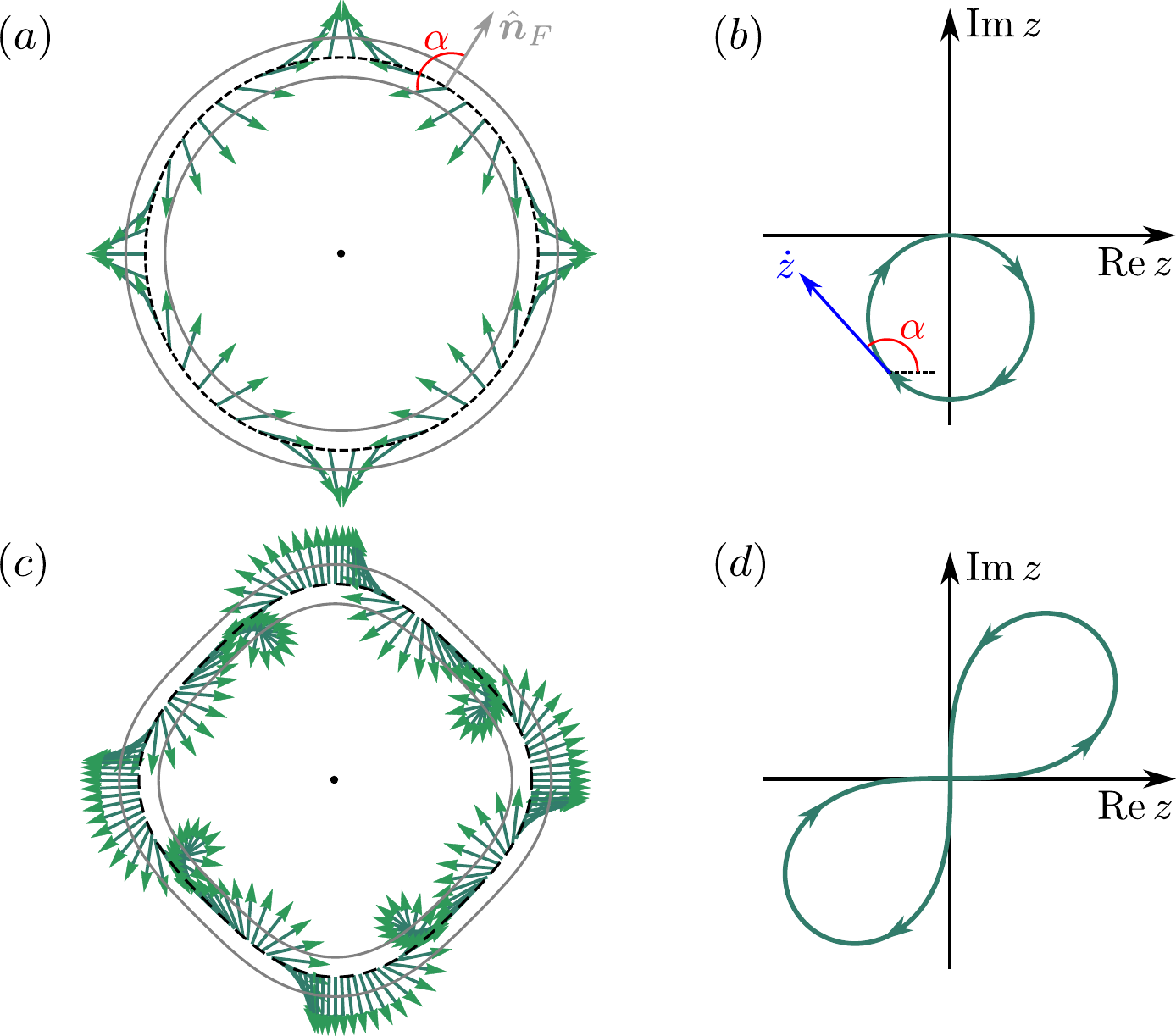}
\caption{Part (b) and (d) show two examples of complex trajectories $z(t)$ that are allowed by \equref{ComplexEquation} with corresponding spin-textures (for $\eta = +1$) given in (a) and (c), respectively. In the limit of a spherical Fermi surface, the texture of (a) coincides with the $l=3$ Dresselhaus coupling in \equref{SpinOrbitVecGen}.}
\label{DifferentSpinOrbitTerms}
\end{center}
\end{figure}

However, spin-orbit coupling in higher angular momentum channels ($l \geq 3$) can be generated spontaneously and if this does occur, some rather exotic behavior follows as we discuss next. Focusing for simplicity on $d=2$, the remaining possible spin-orbit vectors are of the form
\begin{equation}
 \vec{g}(\vec{k}) = g_0 \left( \cos(l\varphi_{\vec{k}}), \pm \sin(l\varphi_{\vec{k}}),0 \right)^T \hspace{-0.2em}, \,\,\, l = 3,5,\dots, \label{SpinOrbitVecGen}
\end{equation}  
with residual symmetry group $\text{SO}(2)_{L_3/l\pm S_3}$ generated by the combination $L_3/l\pm S_3$ of the out-of-plane components $L_3$ and $S_3$ of the orbital and spin angular momentum operators. For the upper (lower) sign, \equref{SpinOrbitVecGen} can be seen as generalizations of the Rashba (Dresselhaus) spin-orbit term with $\vec{g}$ winding $w=l$ ($w=-l$) times on the Fermi surface. The generalized Dresselhaus term with $l=3$ is illustrated in \figref{DifferentSpinOrbitTerms}(a). 

The unconventional form (\ref{SpinOrbitVecGen}) of the spin-orbit coupling has interesting physical consequences: The Berry curvature (finite in the presence of a Zeeman term $\sum_{\vec{k}}\psi^\dagger_{\vec{k}}h\sigma_3\psi^\pdagger_{\vec{k}}$) is enhanced by a factor of $l$ compared to the usual Rashba-Dresselhaus scenario which affects many electronic properties \cite{ReviewBerryPhaseEffects}.
E.g., the anomalous Hall conductivity $\sigma_{xy}$ \cite{Note1}, relating an applied electric field to a perpendicular electric current for $h\neq 0$, is enhanced by a factor of $l$, $\sigma_{xy} = l \cdot  \sigma_{xy}|_{l=1}$.
To illustrate another consequence of a spin-orbit vector of the form (\ref{SpinOrbitVecGen}), let us assume that, in analogy to the proposal of \refcite{SauSemiconductorProp}, the system shows superconductivity in the $s$-wave channel.  Describing the latter on the mean-field level by $\frac{\Delta}{2}\sum_{\vec{k}} \psi_{\vec{k} \uparrow}^\dagger \psi_{-\vec{k} \downarrow}^\dagger + \text{H.c.}$ and focusing on the relevant regime \cite{SauSemiconductorProp} where the chemical potential lies in the Zeeman-induced gap at $\vec{k}=0$, we follow \refcite{AliceaSemicondDiscussion} and project the theory onto the lower, effectively spinless, band. We find that this low-energy model exhibits $(k_1 \mp ik_2)^l$-wave pairing. Recalling that $l \geq 3$, this not only corresponds to very exotic pairing but also leads to a topological class-D invariant $\nu = l$ and, thus, $l$ chiral Majorana modes at the edge of the system.

\section{Lattice effects}

Finally, let us discuss the modifications in the presence of a lattice, i.e., when the periodic potential $U(\vec{r})$ in \equref{eq:Hamiltonian} is finite.
Since this additional term again commutes with the spin and charge density, we can still exclude all phases with order parameter of the form $O_{jj'}=\sum_{\vec{k}} \psi^\dagger_{\vec{k}} \sigma^j k_{j'} \psi^{\pdagger}_{\vec{k}}$, $j=0,1,2,3$, $j'=1,\dots ,d$. Note that this even holds when ions are taken into account as dynamical degrees of freedom since all interactions are functions of the electron and ion density alone.
The main difference is that we cannot rule out any of the irreducible representations of the lattice point group $G_p$ since $k_j$ and $k_jf(\varphi_{\vec{k}})$ transform identically under $G_p$ for any function $f$ that is invariant under $G_p$.    
Nonetheless, our results still lead to significant restrictions. To see this, let us first introduce fermionic operators $c_{n\vec{q}}$, with band index $n$ and crystal momentum $\vec{q}$, diagonalizing the non-interacting part $H_0$ of the Hamiltonian with resulting bandstructure $E_{n\vec{q}}$. Assuming that only one band $n=n_0$ is relevant, it holds $O_{jj'}=\sum_{\vec{q}} c^\dagger_{n_0\vec{q}} \sigma^j v_{j'}(\vec{q}) c^{\pdagger}_{n_0\vec{q}}$, where the summation is restricted to the first Brillouin zone and $v_j(\vec{q}) \equiv \partial_{q_j} E_{n_0\vec{q}}$.
To proceed, we focus on $d=2$ and treat the interaction-induced spin-orbit coupling on the mean-field level, i.e., add $\sum_{\vec{q}} c^\dagger_{n_0\vec{q}} \, \vec{g}(\vec{q})\cdot \vec{\sigma}c^{\pdagger}_{n_0\vec{q}}$ with  $\vec{g}(\vec{q}) = (R_1(\vec{q}),\eta R_2(\vec{q}),0)^T$ to the Hamiltonian where $R_1$ ($R_2$) transforms as $x$ ($y$) under $G_p$ and $\eta = \pm 1$.
Evaluating the constraint $\braket{O_{jj'}} = 0$ in the regime where the Fermi energy is the largest relevant energy scale of the system, we obtain the equivalent condition
\begin{equation}
 \int_0^{\varphi_{\text{ir}}(G_p)} \diff \varphi \, w(\varphi) |\vec{R}(\varphi)|  \, e^{i\alpha(\varphi)}   = 0, \label{ComplexEquation}
\end{equation} 
where $\varphi$ is the polar angle parameterizing the Fermi surface (restricted to the irreducible part of the Brillouin zone, $0<\varphi<\varphi_{\text{ir}}(G_p)$), $w(\varphi) = \vec{k}^2_F(\varphi)/|\hat{\vec{n}}_F(\varphi) \vec{k}_F(\varphi)|$ is a Fermi surface weight function, depending on the Fermi momentum $\vec{k}_F$ and the Fermi-surface normal $\hat{\vec{n}}_F$, and $\alpha(\varphi)$ denotes the angle between $\vec{R}(\varphi)$ and $\hat{\vec{n}}_F(\varphi)$ [see \figref{DifferentSpinOrbitTerms}(a)]. Upon defining $z(t) := \int_0^{t}\diff \varphi \, w(\varphi) |\vec{R}(\varphi)|  \, e^{i\alpha(\varphi)}$, we see that any spontaneously generated spin-orbit texture must lead to a closed trajectory $\{z(t)|0<t<\varphi_{\text{ir}}(G_p)\}$. This restriction is illustrated in \figref{DifferentSpinOrbitTerms}(b-d) for the point group $G_p=C_4$ where $\varphi_{\text{ir}}=\frac{\pi}{2}$ and the boundary condition $\alpha(0) = \alpha(\varphi_{\text{ir}})\,\text{mod}\,2\pi$ is dictated by rotational symmetry.
Most importantly, we see in \figref{DifferentSpinOrbitTerms}(c-d) that, as opposed to the continuum limit $G_p = \text{SO}(2)_L$, spin-orbit vectors with net winding $w = \pm 1$ are possible albeit with much more complex structure than the conventional Rashba or Dresselhaus spin-orbit coupling as dictated by \equref{ComplexEquation}. 

\section{Conclusion}
In summary, we have shown that neither a charge- nor a spin-Pomeranchuk instability with $l=1$ can occur in a non-relativistic \change{metallic} solid-state system.
\change{The divergence of the quasiparticle susceptibility in the $l=1$ spin channel and even the complete quasiparticle contribution is in fact removed by the vanishing of the vertex coupling quasiparticles and real electrons. The actual response may be calculated exactly and is found to be completely non-singular.  The identical result follows within FL theory, if care is taken that the quasiparticle spin current receives a correction term induced by the quasiparticle energy change.}
\change{Our findings imply} that relativistic spin-orbit coupling with residual symmetry group $\text{SO}(3)_{L+S}$ cannot be spontaneously generated. 
Furthermore, any realistic lattice model involving spontaneously generated spin-orbit vectors $\vec{g}(\vec{q}) = (R_1(\vec{q}),\pm R_2(\vec{q}),0)^T$ must satisfy the severe constraint in \equref{ComplexEquation} which is illustrated geometrically in \figref{DifferentSpinOrbitTerms}.

\vspace{1em}

\begin{acknowledgments}
\noindent We acknowledge fruitful discussions with B.~Jeevanesan and S.~Sachdev.
\end{acknowledgments}

\def\theequation{A\arabic{equation}}
\setcounter{equation}{0}


\appendix
\section{Ward identity and its relation to susceptibilities}
\label{Ward identity and its relation to susceptibilities}
In this appendix, we discuss how the Ward identity leading to \equsref{eq:dynsus}{eq:currsus} of the main text is derived.

We first note that if \equref{Prerequ} holds, the dynamics of the density will be governed by the noninteracting
part $H_0$ of the Hamiltonian, $\partial_{t}\rho_{\vec{q}}^{\phi}=i\left[H_{0},\rho_{\vec{q}}^{\phi}\right]$,
and leads to a continuity equation with current $\vec{J}_{\hspace{-0.1em}\phi} = \sum_{\vec{k}\alpha\beta}\frac{\partial\varepsilon_{\vec{k}}}{\partial\vec{k}}\psi_{\vec{k}\alpha}^{\dagger}\phi_{\vec{k}}^{\alpha\beta}\psi^{\pdagger}_{\vec{k}\beta}$.
In order to determine the associated susceptibilities $\chi_\rho$ and $\chi_J$, we analyze the correlator
\begin{align}
\nonumber
 &L_{\vec{k},\vec{k'},\vec{q}}^{\alpha\beta\gamma\delta}\left(\tau,\tau'\right)
 \\
 &=\left\langle T_\tau\,\psi_{\vec{k}+\frac{\vec{q}}{2}\alpha}^{\dagger}(\tau)\psi^{\pdagger}_{\vec{k}-\frac{\vec{q}}{2}\beta}(\tau)\psi_{\vec{k}'-\frac{\vec{q}}{2}\gamma}^{\dagger}(0)\psi^{\pdagger}_{\vec{k}'+\frac{\vec{q}}{2}\delta}(\tau')\right\rangle,
\end{align}
where $\tau$ denotes imaginary time and $T_\tau$ the time-ordering operator. Following \cite{Behn78},
one obtains from the Heisenberg equation of motion of $\rho_{\vec{q}}^{\phi}$
the Ward identity for $L_{\vec{k},\vec{k'},\vec{q}}^{\alpha\beta\gamma\delta}\left(\tau,\tau'\right)$.
After Fourier transformation to frequencies, it reads
\begin{align}
   \nonumber
   \sum_{\vec{k},\alpha\beta}\left(i\Omega-\left(\varepsilon_{\vec{k}+\vec{\frac{q}{2}}}-\varepsilon_{\vec{k}-\frac{\vec{q}}{2}}\right)\right)\phi_{\vec{k}}^{\alpha\beta}L_{\vec{k},\vec{k}',\vec{q}}^{\alpha\beta\gamma\delta}(i\Omega,i\Omega') 
   \\
   =  \phi_{\vec{k}'}^{\gamma\delta}\left(G_{\vec{k}'+\frac{\vec{q}}{2}\delta}\left(i\Omega'\right)-G_{\vec{k}'-\frac{\vec{q}}{2}\gamma}\left(i\Omega'+i\Omega\right)\right), \label{TheWI}
\end{align}
where $G_{\vec{k}}(i\Omega)$ is the exact single-particle Green's function on the imaginary axis. Its retarded analytic continuation has been introduced in \equref{eq:Gtot}. 

The Ward identity (\ref{TheWI}) allows to draw conclusions for the two susceptibilities
\begin{subequations}\begin{align}
\chi_{\rho}(\vec{q},i\Omega)  &=  T \sum_{\vec{k},\vec{k}',\Omega'}\sum_{\alpha\beta\gamma\delta}\phi_{\vec{k}}^{\alpha\beta}\phi_{\vec{k}'}^{\gamma\delta}L_{\vec{k},\vec{k}',\vec{q}}^{\alpha\beta\gamma\delta}(i\Omega,i\Omega'), \\
\chi_{J}^{ij}(\vec{q},i\Omega)  &=  T  \sum_{\vec{k},\vec{k}',\Omega'}\sum_{\alpha\beta\gamma\delta}\phi_{\vec{k}}^{\alpha\beta}\frac{\partial\varepsilon_{\vec{k}}}{\partial k_{i}}\phi_{\vec{k}'}^{\gamma\delta}\frac{\partial\varepsilon_{\vec{k}'}}{\partial k'_{j}}L_{\vec{k},\vec{k}',\vec{q}}^{\alpha\beta\gamma\delta}(i\Omega,i\Omega'). \label{CurrentSusc}
\end{align}\end{subequations}
Multiplying both sides of \equref{TheWI} by $\phi_{\vec{k}'}^{\gamma\delta}$ and $\phi_{\vec{k}'}^{\gamma\delta}\frac{\partial\varepsilon_{\vec{k}'}}{\partial k'_{j}}$, summation over $\vec{k}'$, $\Omega'$ as well as $\gamma$, $\delta$ readily leads (after analytic continuation $i\Omega\rightarrow\omega+i0^{+}$) to \equsref{eq:dynsus}{eq:currsus}, respectively.

\section{Response functions within FL theory}
\label{Response functions within FL theory}
In this section, we first show that the exact result for the spin-current susceptibility resulting from the Ward identity (\ref{TheWI}) can alternatively be obtained within FL theory and then apply this approach to the $l=2$ spin and charge channel.

\subsection{Spin-current susceptibility.} 
To obtain the correct form for the spin current, we use that the distribution function $n_{\vec{k}\sigma}(\vec{r},t)$ obeys the Landau-Boltzmann
equation. Linearized in the external field and Fourier transformed,
$\delta n_{\vec{k}\sigma}(\vec{r},t)\equiv n_{\vec{k}\sigma}(\vec{r},t) - n^0_{\vec{k}\sigma}(\vec{r},t)= \sum_{\vec{q},\omega}\delta n_{\vec{k}\sigma
}(\vec{q},\omega)e^{i\vec{q}\vec{r}-i\omega t}$, the latter reads
\begin{equation}
(\omega-\vec{q\cdot v}_{\vec{k}})\delta n_{\vec{k}\sigma
}+\vec{q\cdot v}_{\vec{k}}\frac{\partial n_{\vec{k}}^{0}}%
{\partial\varepsilon_{\vec{k}}}\delta\varepsilon_{\vec{k}\sigma}=\delta I
\label{qp-LB}%
\end{equation}
where $\delta\varepsilon_{\vec{k}\sigma}=\delta\varepsilon_{\vec{k}\sigma}^{ext}%
+\rho_F^{-1}\sum_{\vec{k}^{\prime},\sigma^{\prime}}F^{\sigma,\sigma'}_{\vec{k},\vec{k}'}\delta n_{\vec{k}^{\prime}\sigma'}$ 
and $\vec{v}_{\vec{k}}=\vec{\nabla}_{\vec{k}}\varepsilon_{\vec{k}%
}=\vec{k}/m^{\ast}$ is the quasiparticle velocity, $\delta I$ is the collision integral, and $n_{\vec{k}}^{0}$ is
the equilibrium (Fermi) distribution function. The applied external field
leads to a quasiparticle energy contribution $\delta\varepsilon_{\vec{k}%
\sigma}^{ext}$. In the case of interest to us (spin-current response)
$\delta\varepsilon_{\vec{k}\sigma}^{ext}=-\frac{1}{m}\sigma\vec{k}\cdot \vec{A}$, where $\sigma\vec{A}$ is a spin-dependent vector potential.

The spin conservation law $\omega\delta n_{S}-\vec{q}\cdot \vec{j}_{S}=0$
follows by multiplying \equref{qp-LB} by $\sigma$ and summing
over $\vec{k},\sigma$, where the collision integral drops out on account of the spin conservation in
two-particle collision processes. The spin density $\delta n_{S}$ and the spin
current density are defined as
\begin{align}
\nonumber
\delta n_{S} &=\sum_{\vec{k},\sigma}\sigma\delta n_{\vec{k}\sigma}, \qquad 
\\
\vec{j}_{S} &=\sum_{\vec{k},\sigma}\sigma\vec{v}_{\vec{k}%
}\left[\delta n_{\vec{k}\sigma}-\frac{\partial n_{\vec{k}}^{0}}%
{\partial\varepsilon_{\vec{k}}}\rho_F^{-1}\sum_{\vec{k}^{\prime},\sigma^{\prime}%
}F_{\vec{k},\vec{k}'}^{\sigma,\sigma'}\delta n_{\vec{k}^{\prime
}\sigma'}\right]. \label{SpinCurrentDensity}
\end{align}
We observe that the spin current consists of two contributions: a ``direct''
term and a ``backflow'' term. Using the usual expansion of the Landau
interaction function $F^{\sigma,\sigma'}_{\vec{k},\vec{k}'}$ (given in the main text) and 
assuming $\delta n_{\vec{k}\sigma}\propto\sigma\vec{k} \cdot \vec{A}$, we have
\begin{align}
\sum_{\vec{k}^{\prime},\sigma^{\prime}}F_{\vec{k},\vec{k}'}^{\sigma,\sigma'}\delta n_{\vec{k}^{\prime}\sigma'}  
 \approx F_{1}^{a} \left(\frac{m^{\ast}}{k_{F}}\right)^{2}\sigma\vec{v}_{\vec{k}}\sum_{\vec{k}^{\prime},\sigma^{\prime}%
}\sigma^{\prime}\vec{v}_{\vec{k}^{\prime}}\delta n_{\vec{k}^{\prime}\sigma'}.
\end{align}
In the limit of $\omega,\vec{q}\rightarrow0$, it holds
\begin{align}
\nonumber
\delta n_{\vec{k}\sigma}    =\delta\left[\frac{1}{\exp\beta
(\varepsilon_{\vec{k}\sigma}+\varepsilon_{\vec{k}\sigma}^{ext})+1}\right] 
\\
= \left(\delta\varepsilon_{\vec{k}\sigma}-\frac{1}{m}\sigma\vec{k}\cdot {A}\right) \frac{\partial n_{\vec{k}}^{0}}{\partial\varepsilon_{\vec{k}}}
\end{align}
with solution
\begin{equation}
\delta n_{\vec{k}\sigma}=-\frac{\partial n_{\vec{k}}^{0}}{\partial\varepsilon_{\vec{k}}}\frac{1}{m}\sigma(\vec{k}\cdot \vec{A})\frac
{1}{1+F_{1}^{a}/3}.
\end{equation}
Substituting into the expression (\ref{SpinCurrentDensity}) for the spin current density, we find
\begin{align}
\nonumber
\vec{j}_{S}   &=\frac{1}{1+F_{1}^{a}/3}\sum_{\vec{k},\sigma}\sigma
^{2}\vec{v}_{\vec{k}}\frac{\vec{k}\cdot \vec{A}}{m}\left(-\frac{\partial
n_{\vec{k}}^{0}}{\partial\varepsilon_{\vec{k}}}\right)\left[1+F_{1}^{a}/3\right]  
\\
&=\frac{1}{3}\frac{k_{F}^{2}}{m^{2}}\frac{m}{m^{\ast}}\rho_{F}
\vec{A} =\frac{n}{m}\vec{A}%
\end{align}
in agreement with the microscopic result following from the Ward identity.


\subsection{The case $l=2$.} 
The phenomenological derivation using the kinetic equation may be extended to
higher angular momentum channels, employing some additional assumptions. We
consider the case $l=2$. To begin with the spin channel, the change in the
quasiparticle distribution function caused by an external field in the spin
channel of $l=2$ symmetry,
\begin{equation}
\delta\varepsilon_{\vec{k}\sigma}^{ext}=-\frac{1}{m}\sigma k_{\alpha}k_{\beta
}\,\delta D_{\alpha\beta}, \qquad \alpha\neq\beta,
\end{equation}
is given by
\begin{equation}
\delta n_{\vec{k}\sigma}=-\frac{\partial n_{\vec{k}}^{0}}{\partial\varepsilon_{\vec{k}}}\frac{1}{m}\sigma k_{\alpha}k_{\beta}\, \delta
D_{\alpha\beta}\frac{1}{1+F_{2}^{a}/5}.
\end{equation}
The quasiparticle distribution is not an observable quantity. A possible
observable is the momentum (charge or spin) current density. Although the
(charge or spin) current is not a conserved quantity, we may still derive an
expression for it from the kinetic equation (\ref{qp-LB}), 
\begin{widetext}
\begin{align}
\omega\vec{j}_{S}  &  =\sum_{\vec{k},\sigma}\sigma\vec{v}%
_{\vec{k}}[\omega\delta n_{\vec{k}\sigma}-\frac{\partial n_{\vec{k}%
}^{0}}{\partial\varepsilon_{\vec{k}}}\rho_F^{-1}\sum_{\vec{k}^{\prime},\sigma^{\prime
}}F^{\sigma,\sigma'}_{\vec{k},\vec{k}'}\omega\delta n_{\vec{k}%
^{\prime}\sigma'}]\nonumber\\
&  =\sum_{\vec{k},\sigma}\sigma\vec{v}_{\vec{k}}\left[\vec{q\cdot
v}_{\vec{k}}\delta n_{\vec{k}\sigma}'+\delta I_{\vec{k}%
\sigma} -\frac{\partial n_{\vec{k}}^{0}}{\partial\varepsilon_{\vec{k}}}%
\rho_F^{-1}\sum_{\vec{k}^{\prime},\sigma^{\prime}}F^{\sigma,\sigma'}_{\vec{k},\vec{k}'}\left(\vec{q\cdot v}_{\vec{k}^{\prime}}\delta n'_{\vec{k}^{\prime}\sigma^{\prime}}+\delta I_{\vec{k}^{\prime}%
\sigma^{\prime}}\right)\right]
\end{align}
\end{widetext}
where we defined $\delta n_{\vec{k}\sigma}^{\prime}=\delta n_{\vec{k}%
\sigma}-\frac{\partial n_{\vec{k}}^{0}}{\partial\varepsilon_{\vec{k}}%
}\delta\varepsilon_{\vec{k}\sigma}$. The equation describes the change of the
local current density by flow out of or into the volume element in terms of
the divergence of the momentum spin current density $\Xi_{\alpha\beta}$ and by relaxation processes,
\begin{equation}
\omega j_{S,\alpha}=\sum_{\beta}\Xi_{\alpha\beta}q_{\beta}-i\Gamma_{\alpha
}j_{S,\alpha},
\end{equation}
where the momentum spin current tensor is defined by
\begin{align}
\nonumber
&\Xi_{\alpha\beta}
\\
&=\sum_{\vec{k},\sigma}\sigma v_{\vec{k}\alpha
}\left(v_{\vec{k}\beta}\delta n_{\vec{k}\sigma}'-\frac{\partial
n_{\vec{k}}^{0}}{\partial\varepsilon_{\vec{k}}}\rho_F^{-1}\sum_{\vec{k}^{\prime
},\sigma^{\prime}}F^{\sigma,\sigma'}_{\vec{k},\vec{k}'}v_{\vec{k}^{\prime}\beta}\delta n_{\vec{k}^{\prime}\sigma^{\prime}%
}'\right).
\end{align}
A possible instability of the system with respect to a deformation of the
Fermi surface of $d$-wave type, as expressed by the external field
$\delta\varepsilon_{\vec{k}\sigma}^{ext}$ defined above, should show up as a
divergence of the susceptibility
\begin{align}
\nonumber
&\chi_{l=2}^a=\frac{\delta\Xi_{\alpha\beta}}{\delta D_{\alpha\beta}}%
\\
&=\sum_{\vec{k},\sigma}\sigma v_{\vec{k}\alpha}\left(v_{\vec{k}\beta}%
\frac{\delta n_{\vec{k}\sigma}'}{\delta D_{\alpha\beta}}%
-\frac{\partial n_{\vec{k}}^{0}}{\partial\varepsilon_{\vec{k}}}%
\rho_F^{-1}\sum_{\vec{k}^{\prime},\sigma^{\prime}}F^{\sigma,\sigma'}_{\vec{k},\vec{k}'}v_{\vec{k}^{\prime}\beta}\frac{\delta
n_{\vec{k}^{\prime}\sigma^{\prime}}^{\prime}}{\delta D_{\alpha\beta}}\right).
\end{align}
We may calculate $\chi_{l=2}^a$ by substituting
\begin{equation}
\delta n_{\vec{k}\sigma}^{\prime}=-\frac{\partial n_{\vec{k}}^{0}%
}{\partial\varepsilon_{\vec{k}}}\frac{1}{m}\sigma k_{\alpha}k_{\beta}\,\delta
D_{\alpha\beta}
\end{equation}
where the factor $1/(1+F_{2}^{a}/5)$ present in $\delta n_{\vec{k}\sigma}$
has dropped out. The resulting expression
\begin{equation}
\chi_{l=2}^a=\frac{1}{5}\frac{k_{F}}{m^{\ast}}\frac{n}{m}(1+F_{1}^{a}/3)
\end{equation}
does not diverge when $1+F_{2}^{a}/5\rightarrow0$ and is actually independent
of $F_{2}^{a}$. The susceptibility does, however, diverge when $m^{\ast
}=1+F_{1}^{s}/3\rightarrow0$ or vanishes at $1+F_{1}^{a}/3=0$, both signaling
an instability of the system.

The analogous derivation for the $l=2$ charge susceptibility gives
\begin{equation}
\chi^s_{l=2}=\frac{1}{5}\frac{k_{F}}{m}\frac{n}{m}.
\end{equation}


\end{document}